\documentstyle[12pt,preprint,morefloats,psfig]{aastex}
\slugcomment{To be submitted to The Astrophysical Journal, May 2002}

\lefthead{Gonz\'alez Delgado et al.}
\righthead{Seyferts 2}

\begin{document}

\title{Is a minor-merger driving the nuclear activity in the Seyfert 2 galaxy NGC 2110?}

\author{Rosa M. Gonz\'alez Delgado}
\affil{Instituto de Astrof\'\i sica de Andaluc\'\i a (CSIC), Apdo. 3004, 18080 Granada, Spain}
\affil{Electronic mail: rosa@iaa.es}

\author{Santiago Arribas}
\affil{Space Telescope Science Institute, 3700 San Martin Drive, Baltimore, MD
21218}
\affil{Electronic mail: arribas@stsci.edu}

\author{Enrique P\'erez}
\affil{Instituto de Astrof\'\i sica de Andaluc\'\i a (CSIC), Apdo. 3004, 18080 Granada, Spain}
\affil{Electronic mail: eperez@iaa.es}

\and

\author{Timothy Heckman}
\affil{Department of Physics \& Astronomy, JHU, Baltimore, MD 21218}
\affil{Adjunct Astronomer at STScI}
\affil{Electronic mail: heckman@pha.jhu.edu}

% The abstract environment prints out the receipt and acceptance dates
% if they are relevant for the journal style.  For the aasms style, they
% will print out as horizontal rules for the editorial staff to type
% on, so long as the author does not include \received and \accepted
% commands.  This should not be done, since \received and \accepted dates
% are not known to the author.

\newpage

\begin{abstract}

We report on a detailed morphological and kinematic study of the isolated non-barred nearby Seyfert 2 galaxy NGC 2110.
We combine Integral Field optical spectroscopy, with long-slit and WFPC2 imaging available 
in the HST archive to investigate the fueling mechanism in this galaxy. Previous 
work (Wilson \& Baldwin 1985) concluded that the kinematic center of the galaxy is displaced $\sim$ 220 pc
from the apparent mass center of the galaxy, and the ionized gas follows a remarkably normal rotation curve. 
Our analysis based on the stellar kinematics, 2D ionized gas velocity field and dispersion velocity, and 
high spatial resolution morphology at V, I and H$\alpha$ reveals that: 1) The kinematic center of NGC 2110
is at the nucleus of the galaxy. 2) The ionized gas is not in pure rotational
motion. 3) The morphology of the 2D distribution of the emission line widths suggests the presence of a minor 
axis galactic outflow. 4) The nucleus is blue-shifted with respect 
to the stellar systemic velocity, suggesting the NLR gas is out-flowing due to the interaction with the radio jet. 
5) The ionized gas is red-shifted $\sim$ 100 km/s over the corresponding rotational motion south of the nucleus, and 
240 km/s with respect to the nuclear stellar systemic velocity. This velocity is coincident with the HI red-shifted 
absorption velocity detected by Gallimore et al (1999). 
We discuss the possibility that the kinematics of the south ionized gas could be perturbed by the collision with a small satellite that 
impacted on NGC 2110 close to the center with a highly inclined orbit. Additional support for this 
interpretation are the radial dust lanes and tidal debris detected in the V un-sharp masked image. 
We suggest that a minor-merger may have driven the nuclear activity in NGC 2110.
 
\end{abstract}

% The different journals have different requirements for keywords.  The
% keywords.apj file, found on aas.org in the pubs/aastex-misc directory,
% contains a list of keywords used with the ApJ and Letters.  These are
% usually assigned by the editor, but authors may include them in their
% manuscripts if they wish.


\newpage

\section{Introduction}

The two most important questions concerning active galactic nuclei (AGN)
 are the fundamental nature of their energy source and the mechanism by which the nuclear 
activity is fueled. Even though it is widely accepted that the energy source originates in the accretion
of mass onto a central super-massive black hole (SMBH), starbursts could also play an important
role in the same galaxies (e.g. Gonz\'alez Delgado 2001; Veilleux 2001, and references therein). In fact, 
many examples have been reported
in which the two phenomena co-exist together. At the high-luminosity regime, ULIRGs
are the best examples (Lutz \& Tacconi 1999). Powerful starbursts are also found 
in the central $\sim$ 100 pc of less luminous AGN, such as Seyfert 2 galaxies 
(Heckman et al 1997; Gonz\'alez Delgado et al 1998), 
and even at smaller spatial scales ($\sim$ few pc), as it is the case of the low luminosity AGN NGC 4303 
(Colina et al 2002). 

The origin of the gas that fuels the central engine and the mechanism that brings
the gas to the very central region are still controversial. Observations of high luminosity AGNs indicate
that QSOs and ULIRGs have been gas-rich mergers (Sanders et al 1988;
Canalizo \& Stockton 2001). However, in low luminosity AGNs, the fueling mechanism remains unclear.
Early investigations in Seyfert galaxies have also suggested that galaxy-interaction could provide
the gas fueling onto the SMBH (e.g. Noguchi 1988; Mihos \& Hernquist 1996). Early research suggested
that there is an excess of companions in Seyfert galaxies (Dahari 1984). However, this result has not been confirmed
more recently (De Robertis, Yee \& Hayhoe 1998; Rafanelli et al 1995). Thus, the galaxy-interaction hypothesis 
can be ruled out as the main fueling mechanism in Seyfert galaxies.

The non-axisymmetric gravitational potential originated by a large scale bar in the host
galaxy has also been suggested as an efficient way to drive the gas to the central region (Shlosman, Begelman \& Frank 1990;
Athanassoula 1992). However, it can be ruled out as a general mechanism because 
an excess of bars in Seyfert galaxies has not been confirmed observationally (Adams 1977; Moles, 
M\'arquez \& P\'erez 1995; Hunt et al 1997; Maiolino, Risaliti \& Salvati 1998). 
Studies based on near-infrared imaging, known to trace the stellar bars better than the optical 
observations, arrive also at contradictory results.
Based on HST (NICMOS) observations Mulchaey \& Regan (1997), Regan \& Mulchaey (1999) and Martini \& Pogge (1999)
find no evidence for an excess of stellar bars in Seyferts. On the opposite, Knapen, Shlosman, \& Peletier (2000)
found an excess in the sample of CfA Seyfert galaxies with respect to a control sample. However, this excess of bars
in CfA Seyfert galaxies disappears when galaxies that are in interaction are excluded from the statistics (M\'arquez
et al 2000).

Minor merger between a  disk galaxy and a nucleated less massive galaxy is also proposed as an 
alternative fueling mechanism (Garc\'\i a-Lorenzo et al 1997; de Robertis et al 1998; Taniguchi 1999). 
In this context, it is interesting to note that some kinematic peculiarities found in NGC 1068 have been 
interpreted by Garc\'\i a-Lorenzo et al (1997 and 1999) as the result of a minor merger event.  
Since most galaxies have satellites
 (Zaritsky et al 1997), it is likely that minor mergers drive the fueling mechanism for most of the isolated
Seyfert galaxies. However, there is not yet observational proof of this scenario.

NGC 2110 is a nearby ($\sim$ 31 Mpc), early type S0, isolated galaxy, showing a Seyfert 2 nucleus
(McClintok et al. 1979; Shuder 1980). This galaxy has been extensively studied at radio, NIR,
optical, UV and X-ray wavelengths. These studies have revealed a number of interesting
properties in this object, making it a good target to investigate the minor-merger hypothesis
as the mechanism that drives the gas to the center and feeds the AGN in NGC 2110. Its main 
properties can be summarized as follow:

\begin{itemize}

\item{It is classified as a narrow-line X-ray galaxy (Bradt et al 1978) due to the strong 
X-ray emission, having a hard X-ray luminosity (2-10 KeV) comparable to those
of Seyfert 1 galaxies (Weaver et al 1995). Soft X-ray emission is also detected at 4 arcsec 
North of the nucleus (Weaver et al 1995).}

\item{ VLA radio continuum maps reveal a symmetrical
jet-like radio emission  extended 4 arcsec along the North-South direction (Ulvestad \& Wilson
1983; Nagar et al 1999). More recent VLBA observations at 3.6 cm also show a 
compact ($\leq$ 1 pc) radio core aligned with the 400 pc scale radio jet (Mundell et al
2000).}

\item{ Ground-based narrow band ([OIII] and H$\alpha$+[NII]) images reveal 
extended emission up to 10 arcsec showing an S-shape morphology in the North-South direction
(Wilson, Baldwin and Ulvestad 1985; Pogge 1989). HST pre-Costar [OIII] and H$\alpha$+[NII] images confirm
the alignment of the optical and radio jet, but only a loose anti-correlation between the brightest
optical and radio knots was found (Mulchaey et al 1994). Ferruit et al (1999) have obtained
HST+FOC UV, and optical spectra of the jet to investigate the mechanism that ionizes
the extended gas. They suggest that the ionized high excitation gas is produced by photoionization by 
fast-shock waves probably generated by the propagation of the radio ejecta into the ISM of the 
host galaxy. Shocks, driven by the radio jet, could be also responsible for the strong 
[FeII] $\lambda$1.2567 $\micron$ nebular emission detected in its nuclear and circumnuclear region
(Storchi-Bergmann et al 1999; Knop et al 2001).}

\item{ From the morphological point of view, the optical continuum (Malkan et al 1998)
and NICMOS/WFPC color maps (Quillen et al 1999) show clear asymmetries in the central region.    
Dust lanes at 1 and 4 arcsec to the west of the nucleus are detected.}

\item{The nuclear stellar population was studied by Gonz\'alez Delgado, Heckman, \& Leitherer (2001)
to look for signatures of starbursts. Its optical continuum (stellar features and spectral energy distribution)
is well explained by an old stellar population with a small contribution from a power-law. Thus, 
no obvious signs of a young (age $\leq$ 1 Gyr) stellar population were found. }  

\item{The kinematics of the circumnuclear ionized gas region was studied by Wilson and Baldwin (1985) 
and Wilson et al (1985). They reported a rotation pattern remarkably similar to
that expected for spiral galaxies. They also found that the
rotation center was not coincident with the optical (and radio)
nucleus, but was located 1.7 arcsec (220 pc) to the south of the nucleus. They also reported the presence
of asymmetric profiles in the [OIII] nebular lines, south and east of the
nucleus.}

\item{HI absorption has been detected in NGC 2110 by Gallimore et al (1999) located south 
of the maximum of the radio continuum source. This 
absorption has a velocity of 2626 km/s; red-shifted  by about 290 km/s with respect to the systemic
velocity, 2335 km/s, calculated by Strauss et al (1992) using the nuclear optical emission lines.} 

\end{itemize}

Here, we present new Integral Field Spectroscopic (IFS) data at optical
wavelengths of the circumnuclear region of NGC 2110. This technique is
ideal for studying an asymmetric and complex object like NGC 2110
since it provides 2D spectral information of relevant spectral features
(i.e., velocity fields, velocity dispersion maps, line intensity maps,
continuum maps, etc) simultaneously and, therefore, without the
uncertainties associated with centering, reference systems, etc.  In
addition to these IFS data, we also present new longslit spectroscopy
to study the stellar kinematics.
These data, together with  some WFPC2
images retrieved from the HST archive, allow us to investigate the
kinematic peculiarities found in NGC 2110 and their possible connection 
with the fueling mechanism of the AGN in NGC 2110. The paper is organized as follows:
In Section 2 we present the observations and data reduction.
Section 3 presents the analysis of the stellar and ionized gas distribution, and section 4
the gas and stellar kinematics. We discuss our results and their implications in section 5. 
The conclusions and summary are in section 6.

\section{Observations and data reduction}

Two types of observations have been carried out: 1) Integral field spectroscopy and 
2) long-slit spectroscopy. In addition, HST WFPC2 images have been retrieved from the 
ST-ECF archive. 

\subsection{Integral field spectroscopy}

NGC 2110 was observed with the INTEGRAL system (Arribas et al 1998) and the Wide 
Field Fiber Optic Spectrograph (Bingham et al 1994) attached to the 4.2m William Herschel 
Telescope at the Roque de los Muchachos Observatory during 1999 February 10. We used the SB2
bundle of fibers, consisting of 219 fibers with a core diameter of
0.9 arcsec each. The 189 data fibers are arranged forming a rectangle with a 16.5$\times$12.3 arcsec
field of view. A ring of 30 fibers, at 45 arcsec from the center, is intended for collecting
the background light. We observed with the R600B grating, 
covering the 3700--7000 \AA\ spectral range with a linear dispersion of 3 \AA/pixel. The mean
effective spectral resolution is 5.3 \AA. The total integration time was 7200 s split into four 
1800 s exposures. The airmass of the object during the observations 
was between 1.25 and 1.45, and the seeing $\sim$ 1.0-1.2 arcsec.

The reduction of the spectra  was performed with the IRAF\footnote{IRAF is distributed by the 
National Optical Astronomy Observatories, which are operated by the Association of Universitites
for Research in Astronomy, Inc., under cooperative agreement with the National Science Foundation}  
data processing software. We followed
the standard procedures applied to spectra obtained with two-dimensional fiber spectrographs 
(Arribas et al 1997). They consist of: 1) Bias subtraction. 2) Cosmic rays rejection. 3) Definition
and extraction of the 219 apertures. 4) Arc lines identification and dispersion correction.
5) Flat-field and throughput correction. 6) Sky subtraction. 7) Atmospheric extinction correction
and flux calibration. The typical wavelength calibration error 0.2-0.3 \AA, gives velocity 
uncertainties of 10-15 km/s. 

To do the flux calibration, we have also observed with the bundle SB2 the spectroscopic standard 
Feige 34. Two 180 s exposures were taken; one of them was taken through a
color filter  that blocks the second-order blue light from wavelengths below 4950 \AA. The 
sensitivity function was built using the stellar spectrum of one of the fibers, then scaling it by 
a factor that represents the fraction of the flux detected by this fiber with respect to the total flux.

\subsection{Analysis of the Integral field data}

Two-dimensional maps of some spectral features, such as continuum, absorption lines equivalent 
width, line emission fluxes, and radial and dispersion velocity were obtained from the fibers individual
spectra with a program that uses a Renka+Cline two-dimensional interpolation method. This routine (NAG-E01SAF) 
transforms a file with the actual position of the fibers in the focal plane of the telescope and 
the spectral feature into a regularly spaced image, and it guarantees that the constructed surface is 
continuous and has continuos first derivatives. The interpolant is then evaluated regularly each 
0.0846 arcsec on a grid of $201\times201$ pixels to create the maps. Note that the continuity conditions imposed to the 
interpolant allows one to evaluate it with a fine sampling. Although this procedure does not modify the 
effective spatial resolution, it improves the general appearance of the maps. 

Prior to interpolation, the kinematical and line intensity 
information for each fiber was obtained measuring the main nebular lines such as H$\beta$,
[OIII] $\lambda$5007, [OI] $\lambda$6300, H$\alpha$, [NII] $\lambda$6584, and 
[SII] $\lambda$6717, 6731. The measurements were done by fitting gaussians to the nebular lines using 
the program $\it{Longslit}$ (Wilkins \& Axon 1991). These fits determine the central wavelength, 
the emission line width, the flux and the continuum. Alternatively, the continuum and emission line images were  
also obtained integrating the fluxes between appropriate wavelength ranges. These images reveal the same 
continuum and nebular emission structures that those obtained by fitting gaussians
to the nebular lines. The image obtained as the difference between the map 
obtained through the gaussian fitting procedure and that obtained integrating the fluxes 
does not show any significant residual emission structure. Thus, this result indicates, 
that the gaussian fitting method works well to provide the continuum and emission line structures of NGC 2110.

 \subsection{Long-slit spectroscopy}

NGC 2110 was observed using the Richey-Chretien spectrograph attached to 4m Mayall Telescope
at the KPNO during 1996 February 15 and 16. The detector was a T2KB CCD chip with a spatial sampling 
of 0.7 arcsec/pixel. The slit width was set to 1.5 arcsec and it was oriented near the paralactic angle,
P.A= 6$^{\circ}$. We used the gratings BL420 and KPC-007 to cover the spectral range, 6600-9100 \AA,
and 3400-5500 \AA, respectively. The linear dispersion is 1.52 \AA/pixel for the red and 1.39 \AA/pixel for
blue spectral range. The effective spectral resolution is $\sim$3 \AA.
One exposure of 1800 s was taken at the red, and 2 exposures of 900 s each at the blue wavelengths.
 The seeing was about 1.5 arcsec.

Data reduction was performed using the FIGARO\footnote{Software from the Starlink Project funded by the 
UK PPARC} data processing software. We followed the standard
procedures applied to long-slit spectra: 1) Bias and flat-field correction. 2) Two-dimensional 
wavelength calibration using ARC2D (Wilkins \& Axon  1991). 3) Atmospheric extinction correction
and flux calibration. Several spectroscopic standards were observed with a 10 arcsec wide slit
during the night. In addition, several velocity template K giant stars were observed for the
kinematic analysis.

\subsection{HST archival data}

We obtained three sets of WFPC2 images of NGC 2110 from the ST-ECF HST archive. The images were 
taken through three different filters: F606W, F919W and FR680P15. 
The filter F606W ($\lambda$=6010.6 \AA, $\delta\lambda$=1497 \AA) (P.I. Malkan) includes emission 
from the strong lines [OIII] $\lambda$5007 and H$\alpha$. The integration time was 500s. The filter 
F791W ($\lambda$=8006 \AA, $\delta\lambda$=1304 \AA) (P.I. Mundell) transmits mainly the red 
continuum of the object, and it is free of contamination from strong emission lines; we refer to it as 
I image. The integration consisted of $2\times120$s exposures. 
FR680P15 (P.I. Mundell) is a linear ramp filter designed to observe the emission line H$\alpha$. 
The band is centered at 6610 \AA, and it has a width of 41 \AA; this filter also includes 
the emission line [NII] $\lambda$6548. The unvigneted field of
view through this filter is about 9 arcsec. The total integration time of 1200 s was splitted 
into four exposures. We refer to it as H$\alpha$. The three sets of observations place the nucleus of NGC 2110 in the PC. 
The central part of NGC 2110 (corresponding to most 
of the emission detected in the galaxy) is thus sampled at 0.046 arcsec/pixel. 

 We use the standard pipeline reduction procedures. The individual frames of the I and H$\alpha$ images were 
combined with {\it crreject} command to eliminate the cosmic rays. To produce a continuum free H$\alpha$ image, 
the I image was scaled down by a factor of 3 and subtracted from the H$\alpha$ plus underlying continuum. 
The scaling of the I image was performed by visual inspection through photometric cuts along the major 
and minor axes. This method was preferred to the direct scaling of the filter widths, because as seen in the 
spectra (see Figure 2) the continuum is quite red, thus, there is a strong color component in the scaling from 
I to the H$\alpha$ wavelength band, that is position dependent.

\section{Stellar and ionized gas distribution}

Figure 1 shows the individual calibrated spectra corresponding to the 189 fibers in the spectral 
ranges which include H$\beta$ and [OIII] $\lambda$4959, 5007 (a) and H$\alpha$ and [NII] 
$\lambda$6548, 6584 (b). Figure 2 shows the nuclear spectrum from the fiber 110 corresponding to the
maximum of the continuum and presumably the AGN, and two circumnuclear spectra 
corresponding to fiber
123 ($\sim$ 3 arcsec north from the nucleus) and fiber 110 ($\sim$ 2 arcsec south from the nucleus).
These two spectra have been chosen to show the north-south asymmetry in the excitation, measured as
the [OIII]/H$\beta$ ratio. In addition to the nebular lines, the spectra show prominent stellar 
absorption lines, such as CaII K, G band and Mgb. 

\subsection{Continuum morphology}

Figure 3a shows in grey scale the I image of NGC 2110 in the PC. Out of the central 2 arcsec,
the red continuum distribution is very smooth and typical of an early type galaxy, with the isophotes 
oriented along P.A=160-170$^{\circ}$. A mosaic image built combining the PC and the three WF cameras
do not show any additional information about low surface brightness continuum in the out-skirt of the galaxy,
that could reflect any distortion of the morphology. However, in the circumnuclear region, NGC 2110 shows
an asymmetry of its inner isophotes in the east-west direction due to the dust lanes that run north-south
between 1 and 2 arcsec west of the nucleus, and trace a clear spiral patern around the nucleus. These dust lane 
structures are shown more clearly in the un-sharp masked I image (Figure 3b). 
Nuclear spiral dust lanes are commonly found in HST observations of active galaxies (Regan \& Mulchaey 1999;
Martini \& Pogge 1999; Pogge \& Martini 2002). They would correspond to spiral arms formed in non-self-gravitating
disk by hydrodynamic instabilities, and they could represent the paths 
of the gas infalling to the center of the galaxy. This image has been obtained by dividing the original I 
image by a 31 pixel median filtered version of the same image. However, the same structures are seen if
the original I image is divided by the model obtained fitting ellipses to the continuum distribution (see below).

To constrain the photometric center, the position angle of the photometric major axis, and the inclination
of the disk of NGC 2110 we have performed an isophotal fit to the I image using the
iraf task ${\it ellipse}$ on the  ${\it stsdas}$ analysis package. No constraints were imposed on the fits, 
thus, the resulting ellipses are all free in the different parameters (center, ellipticity, etc). The fitted
ellipses are shown over the grey scale I image (Figure 3a). The major axis of the fitted ellipses define the 
position angle of the photometric major axis position angle. This position angle is 163.5$^{\circ}$ from North to East.
The inclination of the disk, obtained from the ellipticity of the isophotes, is 42$^{\circ}$. Both 
values are in very good agreement with previously published values (Wilson \& Baldwin 1985). To get the 
photometric center, we have calculated the difference between the center of each fitted ellipse and the 
location of the maximum of the I image. The median residuals are 0.13 arcsec in RA and 0.07 arcsec in Dec.
The largest difference come from the isophotes more affected by the dust lanes; while the outer isophotes 
give smaller residuals. Therefore, the isophotes indicate that the photometric center of the galaxy
is indeed at the maximum of the I band flux.

Figure 4a presents the central emission through the filter F606W. It shows clearly the 
east-west asymmetry at 1-2 arcsec west of the nucleus due to the dust lanes. The inner 2 arcsec isophotes
show a steeper gradient to the west. We have convolved this image with a gaussian (Figure 4b) to
compare it visually with the IFS continuum images. Figure 4c shows the continuum obtained from the observed 
IFS spectra after integrating the fluxes 
between 5320 and 5440 \AA.

\subsection{Ionized gas morphology}

The HST H$\alpha$ continuum-free image is plotted in Figure 5a. It shows a nuclear point source
plus the circumnuclear emission extended along the north-south direction.
It has a bright plume extended 1 arcsec from the nucleus to the north that is aligned with the radio jet.
This plume is narrowest toward the nucleus. At 1 arcsec south, there are two H$\alpha$ clouds that may be
related with the southern radio lobe. However, as it has been previously reported by Mulchaey et al (1994),
the H$\alpha$ brightness distribution of the plume and the two clouds appear to be anti-correlated with the 
maximum of the lobes of the radio jet.
At larger scales, the H$\alpha$ emission shows an S-shaped distribution that extends 5 arcsec to the north 
and 4 arcsec to the south. The north-side of the S-shaped ionized gas morphology follows the inner edge of the dust lanes
distribution seen in the un-sharp masked I Image (Figure 6). However, at the south, the ionized gas is in between
the two dust lanes, but it does not follow so well the curve of the dust lane. Therefore, this large-scale ionized gas
distribution could be reflecting the gas flows towards the nucleus that is photoionized by the AGN.    

The emission line maps of H$\beta$, [OIII] $\lambda$5007, [OI]$\lambda$6300, 
H$\alpha$, [NII] $\lambda$6584 and [SII] $\lambda$6717, 6732 have been also obtained fitting a single 
gaussian profile to the corresponding nebular line in the IFS. 
Figure 5 compares the HST H$\alpha$ image convolved with a gaussian with $\sigma$ equal to the mean seeing
 with the IFS H$\alpha$ and H$\beta$. The extension of the emission, and the morphology of the
large-scale distribution are remarkably equal. However, the inner 1 arcsec gas distribution seen in the HST 
H$\alpha$ image is lost in our ground-based observations due to the seeing.      
 Within an uncertainty of 0.2 arcsec, the peak of the optical continuum is 
coincident with the maximum of the nebular lines. 

As previously reported (Wilson et al 1985,
Pogge 1989, Mulchaey et al 1994), the [OIII] emission  (Figure 7) is more asymmetric than H$\alpha$; it is 
brightest at the north than at the south. The [OIII] emission also extends more than 6 arcsec to the north;
however, at the south, it does not show the larger scale S-shape emission detected in the
 H$\alpha$ and H$\beta$ emission. This high excitation gas is aligned with the radio jet; however, there 
is not a one to one correspondence between the optical and the radio jet (Mulchaey et al 1994).
 The  [OI]$\lambda$6300,  [NII] $\lambda$6584 and [SII] $\lambda$6717, 6732 images also show an
asymmetric morphology, more extended and brighter to the north than to the south. On the north, 
the emission lines show a similar distribution to H$\beta$ and H$\alpha$, but on the south, their
morphology is more similar to the [OIII] emission.  

Excitation maps have been obtained by dividing the emission line images of [OIII] and [NII] by H$\beta$
and H$\alpha$, respectively. These ratios are plotted in Figure 8. There is a clear gradient in 
the [OIII]/H$\beta$ and [NII]/H$\alpha$ ratios. Both images show the lowest ratio 
at the south, between 2 and 4 arcsec from the nucleus. The maximum of the [OIII]/H$\beta$ 
$\sim$ 10, occurs between 3 and 4 arcsec north of the nucleus. A maximum of the [NII]/H$\alpha$,
 $\sim$ 1.6, occurs near the nucleus.  

\subsection{Stellar absorption lines}

In addition to the nebular lines, the circumnuclear spectra show prominent metallic absorption lines, such as CaII
at $\lambda$3933 \AA, G band at $\lambda$4300 \AA, and Mgb $\lambda$5200 \AA. To measure the strength of these
absorption features, we have determined for each individual spectrum a pseudo-continuum in selected
pivot wavelengths, and we have measured the equivalent widths integrating the fluxes
under the pseudo-continuum in defined wavelength windows. We measure the CaII K line in the window
3908-3952 \AA, and the G band in the window 4284-4318 \AA. The pivot-wavelengths that we use are at
3780, 3810, 3910, 4015, 4140, 4260 and 4430 \AA. Once, the equivalent widths are measured, the 
two-dimensional maps are obtained following the method described in section 2.3.

The spatial distribution of the equivalent widths of the metallic lines in the circumnuclear region allows 
us to estimate the possible variation of the stellar population  as a function of the distance to the
nucleus. Thus, relatively low values of the equivalent widths of these lines could be produced by 
the presence of a young stellar population. Alternatively, values of the equivalent widths lower than 
those expected by an old population at the nucleus and circumnuclear region can be caused by scattered
light from the hidden Seyfert 1 nucleus.

Figure 9 shows the iso-equivalent width contours of the CaII K absorption line. The contours are aligned
with the optical jet detected in the emission lines. The strength of this line and of the G band decreases inwards. 
The minimum value of the CaII equivalent width, $\sim$ 11 \AA, occurs nearby the nucleus 
and at 2 arcsec north, where there is also a maximum of the radio and optical jet. This spatial distribution
suggests that extended scattered light from the nucleus could be responsible for the dilution of the
strength of the metallic lines. Alternatively, as it has been previously suggested (Cid Fernandes \& Terlevich
1995; Cid Fernandes et al 2001) the {\em featureless continuum} that dilutes the CaII K line from the old stars in the
host's bulge could be provided by a very young (few Myr old) stellar population.  
 
\section{Ionized gas and stellar kinematics}

\subsection{Velocity field and velocity dispersion map of the ionized gas}

The velocity field was obtained by fitting a single gaussian profile to the emission lines (H$\beta$, 
[OIII] $\lambda$5007, H$\alpha$, [NII] $\lambda$6584 and [SII]$\lambda$6717, 6731). Although the [OIII]
line profile shows evidence for kinematic substructure, the mean velocity field is well represented by the 
centroid wavelength of the brightest component. In fact, in the central 2 arcsec, a blue weaker component
is integrated along the line of sight. This blue component is not clearly detected in H$\alpha$, [NII]
or [SII] in the IFS data, but kinematic structure is clearly detected in the KPNO data. This is due to a 
larger excitation of this second component with respect to the main one, 
combined with the lower spectral resolution of our IFS data.

We have defined the mean nuclear velocity of NGC 2110 as the velocity of the ionized gas in the region where
the maximum of the continuum occurs. This velocity estimated from the H$\beta$ and [NII] lines, 2315 km/s 
(after correcting by the heliocentric velocity of -20 km/s), agrees
well with the systemic velocity reported by Strauss et al (1992) and it is 30 km/s larger than the optical line velocity
reported in the RC3 (de Vaucouleurs et al 1991). It is also in agreement with the nuclear velocity reported by
Wilson et al (1985). 

Figure 10 shows the velocity field of the ionized gas as traced by [NII] and [OIII]. These maps clearly show  
a north-south asymmetry, with a larger velocity gradient in the south than in the north. 
In fact, over 5 arcsec from the optical nucleus along the major
photometric axis, i.e. PA=163$^{\circ}$, the velocities change by about +300
km/s towards the south and about -100 km/s towards the north (see also Figure 11). Despite this result, the
present velocity field does not support the possibility that
the kinematic center of rotation is at 1.7 arcsec towards the south of
the optical nucleus. In fact, the iso-velocity lines crossing that
point are clearly bent towards the reshifted side of the velocity
field and away from the nucleus, far from the expected behaviour for the kinematic minor axis. On
the contrary, the iso-velocity lines crossing the optical nucleus,
though winding, do not show any preferred side, as expected for the true minor kinematic axis.
This clear kinematic asymmetry along the major
photometric axis plus the non-uniform distribution of the blue velocity contours, suggest
that the circumnuclear gas is not normally rotating ambient, interstellar gas in the disk
of the galaxy, as it was previously reported (Wilson et al 1985).

To check the absolute velocity scale and the shape of the velocity field, we have compared
the velocity curve obtained from our KPNO data with a cut on the INTEGRAL velocity field 
along P.A= 6$^{\circ}$ (Figure 12).
We have fitted a gaussian to the KPNO data except in the central 2 arcsec, where two gaussians give a 
better fit. However, in Figure 12 we have plotted only the velocities of the brightest component. The
agreement in amplitude and absolute values between the two curves is very good. Therefore, we are confident
about our estimation of the absolute values of the velocity field within uncertainties of 15 km/s.

The velocity dispersion of the gas has been estimated from the FWHM of the gaussian profile fitted 
to the nebular lines. The INTEGRAL FWHM map has been transformed to a velocity dispersion map after
correcting by the instrumental resolution. Due to the variation in focus of the WYFFOS spectrograph
as a function of the position on the detector, we have obtained an image with the FWHM of the arc
lines spectra measured over the 180 fibers. This image was subtracted quadratically from the FWHM of the
emission line images of NGC 2110.

Figure 13 plots the sigma of the INTEGRAL velocity dispersion of the ionized gas traced by [OIII] and [NII] over 
the photometric isophotes of the continuum at 5000 \AA\ and 6600 \AA, respectively. The velocity dispersion 
shows a bar like morphology oriented along P.A= 60-70$^{\circ}$, which is close to the photometric minor axis.
The maximum of this sigma velocity dispersion, $\sigma$=230 km/s for [NII] and 280-290 km/s for [OIII] (Figure 14), 
is coincident with the peak of the continuum. This result also suggests that the kinematic center is not shifted 
1.7 arcsec to the south of the optical nucleus, but that it is coincident with it. On the other hand, the alignment of 
the iso-dispersion velocity contours with  the minor axis indicates that other non-rotational motions contribute to
the kinematics of the circumnuclear ionized gas. Thus, this gas is not in pure rotational motion. This result
has been observed in edge-on starbursts (Lehnert \& Heckman 1996) and in some Seyferts (Colbert et al 1996).
In both cases, the off-nuclear emission line widths are maximal along the galaxy minor axis. It is believed that 
it is due to a thermal wind driven by the starburst or the AGN.

\subsection{Stellar velocity field and velocity dispersion}

The stellar velocity curve and the velocity dispersion of NGC 2110 along P.A= 6$^{\circ}$ has been inferred 
from the CaII triplet ($\lambda\lambda$ 8498, 8542, 8662) absorption lines via the cross-correlation method 
(Tonry \& Davis 1979). We use the ${\it fxcor}$ task of IRAF between the NGC 2110 spectra (8500-8800 \AA)
and two K giant stars that were observed as velocity templates. 

The shape of the velocity curve obtained with the two stars is very similar. However, the absolute values are 
shifted by 40 km/s. This shift is due to the zero velocity of each standard velocity star being different.
In fact, the CaII lines are blueshifted with respect to the nominal wavelengths in 
the spectrum of one of the stars, and red-shifted by a similar amount (about 0.5 \AA) in the other star.
We take the mean stellar velocity curve.    
We estimate the systemic velocity as the velocity of the maximum of the continuum at 8500 \AA. This velocity, 
2380 km/s (after correcting by the heliocentric velocity of -20 km/s), 
is equal to the velocity that we infer by symmetrizing the stellar velocity curve with respect to 
the nucleus. The amplitude of this curve is $\sim$ $\pm$160 km/s. Typical errors are $\sim$ 30 km/s. 

Figure 15 compares the stellar with the ionized gas (traced by [NII] measured in the IFS  
and H$\alpha$ measured in the KPNO data) velocity curves at P.A= 6$^{\circ}$. 
Note the asymmetry of the velocity curve of the ionized gas with respect to the nucleus, and the larger amplitude 
of this curve with respect to the stellar one. This plot also clearly shows that the ionized nuclear gas is 
blueshifted with respect to the stellar systemic velocity. However, [OIII] nuclear emission blueshifted with respect 
to the stellar motions is commonplace in Seyfert nuclei (e.g. Nelson \& Whittle 1995).

The stellar velocity dispersion along P.A= 6$^{\circ}$ has been also estimated from the width of the 
cross-correlation function. Like the velocity dispersion of the ionized gas, the radial distribution of the 
velocity dispersion also peaks at the position of the nucleus. The stellar velocity dispersion of the nucleus is $\sigma$= 260
$\pm$ 20 km/s, in agreement with the value ($\sigma$=220 $\pm$ 25 km/s) estimated by Nelson \& Whittle
(1995) using the Mgb lines.

\subsection{Kinematics of the nuclear ionized gas}

The kinematics of the ionized gas in the central 2 arcsec is complex. The [OIII] line shows a very asymmetric 
profile with a blue wing that extends up to 2 arcsec to the south from the continuum peak. This result was previously 
reported by Wilson et al (1985). This kinematic structure is fainter in H$\beta$, but is also detected in our KPNO
data in the H$\alpha$, [NII] and [SII] lines. At the nucleus (the maximum of the continuum), H$\alpha$ and [NII]
show two velocity components, with the weaker component at redder wavelengths than the brighter one. However,
at 2 arcsec south of the nucleus, the weaker component is at bluer wavelengths (Figure 16). We have fitted two
gaussians to each line (H$\alpha$ and [NII]$\lambda$6584,6548) to fit the whole profile, and we find that 
the difference in velocity between these two kinematic components is 250 km/s. 
We have plotted these two values in the H$\alpha$ velocity curve
(Figure 15) as blue and red circled dots. Note that these components are blue and red shifted with respect 
to their corresponding radial stellar velocity, suggesting that the central ionized gas is an expanding 
jet or bubble-like kinematic structure, thus implying that the central gas is out-flowing. This 
outflow could be caused by the interaction of the radio jet with the circumnuclear gas.      

We have visually inspected some [OIII] and H$\alpha$+[NII] high resolution STIS spectra (gratings G430M and 
G750M) that are available in the HST archive (P.I. Whittle and Mundell, respectively). Both data sets have been taken 
close to the kinematic major axis with a slit width of 0.2 arcsec. These spectra corroborate our results but they also 
show that the kinematics of the ionized gas at 0.1 arcsec scales is even more complex in detail than
provided by the data we present here.

\section{Discussion}

\subsection{Evidence for non-rotational motions in NGC 2110}

The ionized gas velocity field in NGC 2110 was obtained by Wilson and
Baldwin (1985) on the basis of long-slit spectroscopic observations.
They reported a pattern remarkably similar to the rotation curves
commonly obtained for spiral galaxies, suggesting that the
ionized gas in NGC 2110 was indicating rotation. However, they also
reported an interesting asymmetry. Taking the nucleus as reference, the
velocities increase more rapidly to the south than to the north.  This
result led them to suggest that the true kinematic center (i.e. center
of rotation) was 1.7 $\pm$ 0.4 arcsec towards the south of the optical
nucleus. They proposed several alternatives to explain the offset of the 
kinematic center: a) dust obscuration that hides the true continuum nucleus;
b) the ionized gas is not in gravitational motion; c) stars and gas are orbiting 
in a gravitational potential of a large amount of obscured or dark matter.
They dismissed hypothesis (a) because the outer isophotes of the stellar light
are not centered on their estimation of the kinematic center. Arguments against  
hypothesis (b) and in favor of (c) are the description of their velocity curve in terms 
of a rotation disk. However, for hypothesis (c) to be correct, the interior mass of  
the nucleus should be $\sim$ 8$\times$10$^9$ M$_\odot$, which is too large to 
not let a strong signature on the stellar light isophotes. Therefore, they did not find
any satisfactory interpretation for the velocity field. Our results
indicate that the kinematic center of NGC 2110 is not offset with respect to 
the optical nucleus, and their hypothesis (b) is correct as we have shown in 
section 4 and as we further discuss below.

The most clear indication that the warm gas is not just rotating under the
galaxy gravitational potential comes from the comparison between its
velocity and the stellar velocity along the P.A= 6$^{\circ}$ (Figure 15).
From this figure we infer
that: i) the nucleus is indeed at a symmetric position with respect to
the stellar kinematics, and ii) the amplitude of the ionized gas velocity
curve is larger by 70-90 km/s than the corresponding to the stellar curve. 
The ionized gas kinematics also support this result.
Figures 10 and 13 show the velocity field and the velocity dispersion over NGC 2110 as derived
from gaussian fits to the H$\alpha$ and [NII] emission lines. 
These maps clearly show the asymmetry reported by Wilson \& Baldwin (1985).
However, they  reveal some features which do not suggest rotation.
The asymmetry between the North and South
regions of the velocity field is deeper than just the asymmetry in
the rotation curve. In the red part the maximum is clearly defined at
about 4 arcsec. It is slightly (but significantly) offset from the
major photometric axis. The highest velocities in the blue part (North) are
found in two regions, each at one side of the photometric major axis.
In addition, 'everywhere' in the receding part of the velocity field the
velocities change more rapidly than in the corresponding blue region. 
Moreover, the amplitude of this velocity field ($\sim$ 400 km/s) is very
large if compared with what is typical for early type galaxies (see,
for instance Simien $\&$ Prugniel 2000).
In addition, the gas velocity dispersion map is elongated along the minor 
photometric angle, that  presumably points to the rotation axis of the galaxy.  
 All this suggests that the motions 
of the ionized gas do not indicate gravitationally bound orbits, 
as previously suggested.

To quantify the differences between the ionized gas velocity field with
respect to a pure rotation disk, we have built a 2D rotational model using a 1D
model fitted to the stellar velocity field. We use equation (2-54c) 
of Binney and Tremaine (1987):

\begin{equation}
v_c = \frac{v_0 r}{\sqrt{r^2+r_0^2}}
\end{equation}

\noindent
where $v_0$ is the asymptotic rotational velocity, and $r_0$ the distance from the 
kinematic center at which the velocity arises to $v_0/\sqrt{2}$. 
From a fit to the deprojected stellar velocity curve, we obtain the values 
of $v_0=245$ km/s and $r_0=1.85$ arcsec. With these parameters
we have built the 2D rotational model, projected to the plane of the sky
with an inclination of 42$^{\circ}$, the line of nodes at PA=163$^{\circ}$
and a systemic velocity equal to 2380 km/s (as determined by the stellar rotational curve). 
The predicted velocity field
and the residual map resulting from the subtraction of this rotational disk to the
IFS velocity field are plotted in Figure 17. The residual map shows that the
nucleus is blue-shifted with respect to the systemic velocity, and there is an excess
velocity of up to +100 km/s on the south. The residuals on the north are smaller, less
than 30 km/s. Therefore, we have found strong evidence that the ionized gas in the south of NGC 2110 
is not following the pure gravitational potential of the galaxy.

\subsection{Origin of the out-flowing nuclear ionized gas}

The residual velocity map and the kinematic structure found in the strongest nebular lines
and the two velocity components in H$\alpha$ and [NII] indicate that the NLR gas is flowing out  
of NGC 2110. Additional evidence of an out-flowing gas is the alignment of the maximum of the emission-line widths
and the galaxy minor axis. Other examples of outflows in AGNs have been reported previously. In some cases,
the outflow is spatially resolved at large scales (e.g. Cecil, Bland \& Tully 1990; Colbert et al 1996;
Veilleux, Shopbell \& Miller 2001; Arribas, Colina \& Clements 2001), and in other cases, 
the outflow is not resolved and it is restricted to the NLR (e.g. Whittle 1985). 
Several mechanisms can be responsible for the outflow. They are: 1) a thermal wind from a circumnuclear starburst 
(e.g. Colina, Arribas \& Borne 1999);  2) a thermal wind by the AGN torus (Krolik \& Begelman 1986);  and 
3) collimation by a radio jet (Taylor, Dyson \& Axon 1992) or entrainment along the radio jet. 

We can discard the circumnuclear
starburst model as the origin of the outflow in NGC 2110 based on our previous analysis of the 
nuclear stellar population. The optical continuum distribution and the stellar features do not show any
strong evidence of a circumnuclear starburst, except that a small fraction of the optical light 
could be provided by a young stellar population or by a scattered light from the hidden Seyfert 1 nucleus.
Due to the degeneracy in the slope between a power law and a young ($\leq$ 10 Myr) stellar population
(see Cid Fernandes \& Terlevich 1995) we can not distinguish between these two components. The analysis
of the equivalent width of the stellar features (CaII K and G band) presented here (see Figure 9) also
indicates that the scattered light (or the young stellar component) has to be extended toward the north
of the nucleus. However, in absence of a strong evidence for the existence of a starburst, such as
Wolf-Rayet features, high-order Balmer lines in absorption, etc, we interpret this dilution as produced 
by AGN scattered light.

A detailed analysis of the STIS spectra should distinguish between the other three mechanisms. However,
our analysis and the radio morphology of NGC 2110 points out to the collimated jet-driven outflow
model as the more plausible mechanism for the nuclear outflow. 
The radio continuum observations (Ulvestad \& Wilson 1983; Nagar et al 1999) have revealed
a linear radio structure aligned $\pm$2 arcsec North-South from the nucleus. Mulchaey et al (1994)
have found that the optical emission lines and radio continuum emission extend along similar directions;
however, their brightness distributions appear anticorrelated, and the optical jet more curved than
the radio jet. Pringle et al (1999) have derived the orientation of the radio jet. They found that the jet
is at 50$^{\circ}$ from the line of sight and 29$^{\circ}$ with respect to the line of nodes. Therefore,
the radio jet is not parallel to the rotational axis. This misalignment suggests that the angular momentum
vector of the accretion disk collimating the jet is also misaligned with respect to the host galaxy. 
However, these angles indicate that the radio jet is pointing toward us and, therefore, it explains the 
blueshifted velocities observed in the nuclear region as due to a jet induced outflow.

\subsection{Origin of the non-rotational motions on the south of the galaxy}

Figure 15 indicates that, excluding the nucleus, the stellar and ionized gas velocities fields are in 
very good agreement towards the North. Towards the South, the disagreement is clear, though the residual 
velocities (Figure 17b) do not correlate with the ionized gas distribution (Figure 5). Therefore, 
although the radio jet may play a role to explain the north-south asymmetry in the velocity field of the 
ionized gas, other mechanism is required to explain why the ionized gas is red-shifted by more than about 100 km/s
with respect to the stellar rotation velocity field at the south of the galaxy, and about 240 km/s with respect 
to the nuclear stellar velocity. 

Other evidence that the velocity field at the south is peculiar has been reported by Gallimore et al (1999). 
HI observations with the VLA have detected an HI cloud in absorption, located south of the maximum of the radio 
continuum emission. This cloud has a velocity of 2626 km/s and, therefore, it is red-shifted about 246 km/s with 
respect to the nuclear stellar systemic velocity. Gallimore et al have interpreted this HI absorption 
 as a rotational disk of HI. They argue that we may not see the HI disk in the North
because the North radio jet is in front of the disk as seen by us; at the South, the radio jet is 
behind the disk illuminating it. However, this HI velocity
agrees very well with the velocity of the ionized gas that we measure at the south of the galaxy, both thus 
having an excess +100 km/s with respect to the pure rotation at that radius. Therefore, an alternative explanation 
is that an HI cloud is infalling toward the galaxy and it is impacting close to the center of NGC 2110.

The decoupling between the ionized gas and the stellar velocities could be explained if a minor 
merger is occurring in NGC 2110. The fact that the morphology at large scales is not perturbed 
(we don't see distortion of the isophotal contours or evidence of tidal tails), indicates that 
it would have to be a minor-merger with a small galaxy or an intergalactic HI cloud, that does not 
produce a strong effect on the large scale morphology of the galaxy.
This is supported by previous studies of the effects of a minor merger in a main galaxy.
From a detailed study of a sample of nearby Sa galaxies, Haynes et al (2000) and Haynes (2001) 
find that while these galaxies show no morphological signs of interaction, they do show a range 
of clear kinematical peculiarities that can be interpreted in terms of a kinematic memory of a 
past minor merger.
A similar conclusion is also bore by numerical simulations of minor merger accretion by early
type galaxies (Bak 2000); Bak concludes that there is no major photometric distinction between 
galaxies which have undergone a minor merger and those which have not.

To investigate the minor-merger hypothesis as the origin of the red-shifted HI absorption line 
and the excess of the ionized gas red-shifted velocity at the south of the nucleus, we have further 
inspected the HST images to look for fine detail morphological peculiarities in the center of the galaxy. 
An un-sharp masked F606W image (Figure 18) reveals some morphological
features that have not been detected previously in the I and H$\alpha$ images. 
Besides the dust lanes that trace a spiral pattern toward the nucleus which are common in HST images of galaxies
(e.g. Pogge \& Martini 2002), there are some residual arcs in emission at a radius of 6 to 8 arcsec 
located south to north-east to north of the nucleus. 
In addition, there are also some {\em radial} dust lanes. Note that while the more typical spiraling
dust lanes are located mostly to the north-west side (and south at small radii), 
these {\em radial} dust lanes are mostly located at the south-east to south-west quadrants (at larger
radii), in particular, the most prominent of these lanes is at the south-west pointing directly to the HI cloud. 

These features are very faint and they do not affect the large scale symmetry in
the isophotal analysis, in agreement with the findings of Haynes (2001) and Bak (2000). 
A possible interpretation is that these faint features and kinematic peculiarities in the south 
of the galaxy are produced by a minor-merger with a small satellite or intergalactic HI cloud that passed 
through in a highly inclined orbit close to the nucleus of NGC 2110. Merger or passage of larger 
satellite galaxies do produce similar spoke-like structures, but at significantly larger
scales, that are very clearly reflected on the overall morphology of the galaxy (such as, e.g.,
the Cartwheel). Very detailed analysis of faint structures are needed to find more such 
little mass encounter candidates.

Thus, if the interpretation of these morphological and kinematic peculiarities is correct, we are 
observing a case in which the nuclear activity could be driven by a minor merger. 
Taniguchi (1999) has proposed that the fueling mechanism of isolated non-barred Seyfert galaxies could be due to a minor-merger
between a disk galaxy and a nucleated satellite. If the major galaxy has a gas-poor disk, then, Seyfert activity without 
circumnuclear star formation is triggered. Even more, he suggests that the plane of the accretion disk that forms is 
parallel to the galaxy rotation
disk if the satellite impacts the disk with an almost aligned orbit. On the contrary, when the satellite 
impacts almost perpendicular to the disk, a Seyfert 2 can form with the plane of the accretion
disk highly inclined with respect to the rotational disk of the host galaxy. 
This scenario gives also a plausible explanation for the
misalignment between the orientation of the radio jet and the galaxy rotation axis of NGC 2110. 

Thus, we suggest that the morphological and kinematic peculiarities seen in the south of NGC 2110 could be produced
by a minor-merger of the galaxy with a nucleated satellite that impacts the disk of NGC 2110 close to its nucleus with a 
highly inclined orbit. 
We speculate that this could be the mechanism responsible of triggering the Seyfert activity 
in this isolated non-barred Sa galaxy, that drive the gas toward the center to feed the AGN.

\section{Summary and conclusions}

We have obtained Integral field optical spectroscopy in the range 3700-7000 \AA, and long-slit optical spectroscopy (3400-5500 \AA\
and 6600-9100 \AA) along P.A= 6$^{\circ}$ of the Seyfert 2 galaxy NGC 2110. These observations allow us to get the distributions 
of the emission
lines (H$\alpha$, H$\beta$, [OIII], [OI], [NII], and [SII]), the ionized gas velocity field,
the velocity dispersion map, and the stellar velocity curve along P.A=6$^{\circ}$. These data combined with HST+WFPC2 
(I, F606W and H$\alpha$) have reported the following results:

\begin{itemize}

\item{At large scales, the morphology of NGC 2110 is not perturbed. The red continuum is well fitted by ellipses that have their 
major axis oriented at P.A=163.5$^{\circ}$, and an inclination of  42$^{\circ}$. The centers of these ellipses
are located at the maximum of the continuum within 0.13 arcsec in RA and 0.07 arcsec in Dec. This result
indicates that the photometric center is at the nucleus of the galaxy. }

\item{V and I un-sharp masked images have revealed dust lanes that spiral toward the nucleus and they may represent the
gas flows that feed the AGN. The V (F606W) un-sharp masked image shows also radial dust lanes and emission arcs at 
5 and 8 arcsec at the south and north-east, similar to the tidal debris predicted by numerical simulations 
of the merger of a small galaxy with a galaxy disk.}

\item{The ionized gas is extended in the North-South direction, such as the linear radio jet. The morphology of 
[OI], [NII] and [SII] is similar to [OIII], that is more extended to the north than to the south. This distribution
suggests that the gas at the north may be photoionized by the AGN. At larger scales, H$\alpha$ and H$\beta$ images show 
an S-shape that is similarly extended at the north and at the south.}

\item{The ionized velocity field shows clearly a north-south asymmetry, with a larger velocity gradient in the south 
than in the north. This asymmetry is seen also along the kinematic major axis. This result suggests that the ionized
gas is not normally rotating gas in the disk of the galaxy. The ionized [NII] nuclear gas velocity is 2315 km/s.}

\item{The ionized gas velocity dispersion map shows a bar like morphology oriented near the photometric minor axis. The
maximum of the velocity dispersion is also coincident with the maximum of the optical continuum. Thus, the kinematic 
center is coincident with the nucleus of NGC 2110. This alignment between the maximum of the emission line widths and the
galaxy minor axis may be due to a wind driven by the AGN.}

\item{The stellar velocity curve inferred from the CaII triplet observed along P.A=6$^{\circ}$ is symmetric with respect 
to the nucleus that has a systemic velocity of 2380 km/s. The amplitud of the stellar velocity curve is lower than the amplitud 
of the ionized gas velocity curve.}

\item{The nuclear ionized gas of NGC 2110 is blueshifted with respect to the stellar systemic velocity. 
Two velocity components that are shifted $\pm125$ km/s from the stellar rotation curve have been detected 
at the central 2 arcsec. These components indicate that the 
NLR is out-flowing, possibly driven by the radio jet.}

\item{At larger spatial scales (up to 5 arcsec),  the ionized gas on the south of NGC 2110 is red-shifted by about 100 km/s
with respect to the stellar velocity field, and 240 km/s with respect to the nuclear stellar systemic velocity. 
This excess of the velocity with respect to the rotation indicates that
the gas in the south of NGC 2110 is not following the pure gravitational potential of the galaxy.
It is coincident with the HI red-shifted (240 km/s) absorption extended to the south of the nucleus 
detected by Gallimore et al (1999). }

\end{itemize}

We conclude that the ionized gas on the south is kinematically perturbed. This conclusion together with the morphological peculiarities
detected (arcs, radial dust lanes)
suggests that the nuclear activity could be the result of a minor-merger. The collision with a nucleated small satellite
(or with an intergalactic HI cloud) with a very
high inclined orbit that impacts close to the nucleus may be responsible of the red-shifted velocity of the ionized
gas in the south of the galaxy and the arcs and radial dust lanes observed in the central host disk of the galaxy.
This interpretation of the data is of relevance in relation with the fueling mechanism and the triggering of
nuclear activity in isolated non-barred Seyfert galaxies.

{\bf Acknowledgments}

We are grateful to Jack Gallimore, the referee, for his suggestions and corrections that helped 
to inprove the paper. We also thanks to Luis Cuesta for making his {\it inter2r} program available 
that allows to built the two-dimensional images, Jaime Perea for SIPL, and Luis Colina for his 
comments from a thorough reading of the paper. 
This work was supported by the Spanish DGICYT project AYA2001-3939-C03-01.

The 4.2m William Herschel Telescope is operated by the Isaac Newton Group at the Observatorio de 
Roque de los Muchachos of the Instituto de Astrof\'\i sica de Canarias. 

Based observations made with the NASA/ESA Hubble Space Telescope, obtained from the data archive
at ST-ECF.

% And finally, we must deal with the figures.  There are three figures
% associated with this manuscript; two figures are Encapsulated
% PostScript (EPS) files.  The third figure is a grey scale figure that does
% not exist in EPS form.
%
% Authors have three options for including figure information within a
% manuscript.  Not all the options may be acceptable by the target Journal - be
% sure to look at the appropriate submission instructions, electronicor
% otherwise.
%
% Option 1.  Using this option, only the figure captions are included in the
% main body of the manuscript.  The figure captions must start on a new page.
% The captions are generated with the \figcaption[]{} command: the first
% argument is optional, if you put something in there, put the name of the
% EPS file that goes with the caption; the second argument is the figure
% caption itself, and may include a \label command.  The \figcaption command
% generates the figure numbers.  This option is acceptable for all manuscript
% submissions.

\clearpage

%Figure 1
\begin{figure} 
%\centerline{\psfig{figure=n2110-f1a.ps,width=10cm}} 
%\centerline{\psfig{figure=n2110-f1b.ps,width=10cm}} 

\figcaption{Final calibrated circumnuclear (16.5$\times$12.3 arcsec) spectra of NGC 2110 
in the spectral ranges: (a)  $\lambda$4870-5080 including H$\beta$, [OIII] $\lambda$ 4959, 5007; 
(b) $\lambda$6650-6685 including H$\alpha$, [NII] $\lambda$ 6548, 6584. The numbers in the plot 
indicate the number of the fiber at the entrance of the spectrograph. The orientation is North 
up and East left. The 180 spectra are autoscaled. }

\end{figure} 

%Figure 2
\begin{figure} 
%\centerline{\psfig{figure=n2110-f2.ps,width=10cm}} 

\figcaption{Nuclear and two circumnuclear  spectra of NGC 2110, corresponding to the fibers
110 (nucleus), 100 ($\sim2$ arcsec south of the nucleus) and 123 ($\sim3$ arcsec north of the nucleus).
The circumnuclear spectra are shifted by 2 and 2.3$\times$10$^{-15}$ erg s$^{-1}$ cm$^{-2}$ \AA$^{-1}$. Note the 
strong gradient along the excitation in the north-south direction. }

\end{figure} 

%Figure 3
\begin{figure} 
%%\centerline{\psfig{figure=n2110-f3.cps,width=17cm,angle=270}} 

\figcaption{HST+WFPC2 (PC) images of NGC 2110: a) In grey scale is the continuum emission
through the filter F791W. Over-plotted as full lines are the isophotal contours
fitted to this image. b) Unsharp masked image. Dust lanes trace a spiral patern around the nucleus 
to the nucleus are clearly seen 
at 1 and 2 arcsec to the west. North is up and East to the left. This is the orientation in all the images. }

\end{figure}

%Figure 4
\begin{figure} 
%%\centerline{\psfig{figure=n2110-f4.ps,width=17cm,angle=270}} 

\figcaption{ 
a) HST+WFPC2 image of the central light distribution of NGC 2110 through the filter F606W.
b) F606W image convolved with a gaussian with a $\sigma$=mean seeing. c) IFS
image of the stellar light distribution in the central region of NGC 2110 
traced by a continuum window centered at  5380 \AA.}

\end{figure} 

%Figure 5
\begin{figure} 
%%\centerline{\psfig{figure=n2110-f5.ps,width=17cm,angle=270}} 

\figcaption{ a) HST H$\alpha$ image. b) HST-H$\alpha$ image convolved with a gaussian 
of fwhm equal to the mean of our IFS data. c) IFS H$\alpha$ image. d) IFS H$\beta$ image.}

\end{figure} 

%Figure 6
\begin{figure} 
%%\centerline{\psfig{figure=n2110-f6.ps,width=12cm,angle=270}} 

\figcaption{The un-sharp masked I image is plotted in grey scale. The contours correspond to the
HST-H$\alpha$ image.} 

\end{figure}

%Figure 7
\begin{figure} 
%%\centerline{\psfig{figure=n2110-f7.ps,width=17cm,angle=270}} 

\figcaption{INTEGRAL images of the ionized gas circumnuclear region of NGC 2110 as traced by
(a) [OIII] $\lambda$5007, (b) [OI] $\lambda$6300, (c) [NII] $\lambda$6584, 
(d) [SII] $\lambda$6717, 6731. The contours correspond to: 
(a) 3, 5, 10,15, 20, 40, 60, 80, 100, 150, 250 and 300 $\times$ 10$^{-15}$ erg s$^{-1}$ cm$^{-2}$; 
(b) 3, 5, 15, 25, 50, 75 and 100 $\times$ 10$^{-15}$ erg s$^{-1}$ cm$^{-2}$;
(c) 3, 5, 10, 20, 30, 40, 50, 70, 100, 150, 200, 250 and 300 $\times$ 10$^{-15}$ erg s$^{-1}$ cm$^{-2}$;
(d) 3, 5, 10, 15, 25, 50, 75 and 100 $\times$ 10$^{-15}$ erg s$^{-1}$ cm$^{-2}$.
}

\end{figure} 

%Figure 8
\begin{figure} 
%\centerline{\psfig{figure=n2110-f8.cps,width=17cm,angle=270}} 

\figcaption{INTEGRAL images of the excitation of the 
ionized gas in the circumnuclear region of NGC 2110 as traced by
(a) [OIII] $\lambda$5007/H$\beta$, (b) [NII] $\lambda$6584/H$\alpha$.
Contours correspond to: (a) 1.5 2 3 4 5 6 7 8 9 10 11; 
(b) 0.7 0.8 0.9 1 1.1 1.2 1.3 1.4 1.5 1.6. }

\end{figure} 

%Figure 9
\begin{figure} 
%\centerline{\psfig{figure=n2110-f9.cps,width=10cm,angle=270}} 

\figcaption{Iso-contours (full line) of the equivalent width of the absorption line 
CaII K measured in the individual spectra over the H$\beta$ emission line distribution (dotted line). 
The contours plotted have equivalent width of 11.2, 11.5, 12, 12.5, 13, 13.5 \AA.
The equivalent width decreases inwards,  the inner countor corresponds to 11.2 \AA. }

\end{figure}

%Figure 10
\begin{figure} 
%\centerline{\psfig{figure=n2110-f10.cps,width=17cm,angle=270}} 

\figcaption{Velocity field of the ionized gas in the central region 
of NGC 2110 inferred from a single gaussian profile fit to the emission lines (a) [OIII] $\lambda$5007 and (b)
[NII] $\lambda$6584. Continuum light is also plotted by gray scale and by contours in green (dotted line). 
Iso-velocity lines are plotted in steps of 25 km/s. Blue (dashed line) contours are velocity values below 2315 km/s, 
and red (full line) are velocity values larger than 2315 km/s. The iso-velocity contour of 2315 km/s is plotted by a white 
full line.}

\end{figure} 

%Figure 11
\begin{figure} 
%\centerline{\psfig{figure=n2110-f11.cps,width=15cm,angle=270}} 

\figcaption{Velocity curve of the ionized gas traced by H$\beta$ (dotted green line), 
[OIII] (full blue line) and [NII] (dashed red line) obtained from cuts in the INTEGRAL 
velocity fields along the major (left) and minor (right) photometric angle.}

\end{figure} 

%Figure 12
\begin{figure} 
%\centerline{\psfig{figure=n2110-f12.cps,width=15cm,angle=270}} 

\figcaption{Comparison of the velocity curve of the ionized gas obtained from the KPNO data (red cross) and a cut 
on the IFS velocity field along P.A= 6$^{\circ}$  (blue line). Velocity curves for [OIII] are on the left and 
on the right is for [NII]. }

\end{figure}

%Figure 13
\begin{figure} 
%\centerline{\psfig{figure=n2110-f13.cps,width=17cm,angle=270}} 

\figcaption{Velocity dispersion  of the ionized gas in the central region 
of NGC 2110 inferred from a single gaussian profile fit to the emission lines (a) [OIII] $\lambda$5007 and 
(b) [NII] $\lambda$6584. Continuum light is also plotted by greyscale and by contours in green (dotted line). 
 Iso-sigma velocity dispersion is plotted in steps of
20 km/s with the minimuum contour of 100 km/s, and the maximuum of 340 and 240 km/s for [OIII] and [NII], 
respectively.}

\end{figure} 

%Figure 14
\begin{figure} 
%\centerline{\psfig{figure=n2110-f14.cps,width=15cm,angle=270}} 

\figcaption{As Figure 11 for the ionized gas velocity dispersion.}

\end{figure}

%Figure 15
\begin{figure} 
%\centerline{\psfig{figure=n2110-f15.cps,width=12cm,angle=270}} 

\figcaption{Comparison of the stellar velocity curve (orange stars) along the PA=6$^{\circ}$ inferred
from the cross-correlation in the spectral range of the CaII triplet  with the velocity curve of the ionized
gas traced by the emission line [NII] (green dashed line) inferred from a cut on the velocity 
field map (figure 7a) along PA=6$^{\circ}$ and the velocity curve of H$\alpha$ (dots and circled dots) 
obtained from the KPNO data. The circled dots represent the two kinematic components fitted to the H$\alpha$ line. }

\end{figure}

%Figure 16
\begin{figure} 
%\centerline{\psfig{figure=n2110-f16.cps,width=10cm,angle=270}} 

\figcaption{Profile of the H$\alpha$, and [NII]$\lambda$6548,6584 lines in the KPNO data for the crossection
where the continuum is maximum (nucleus) and the three next crossections to the south. The spatial sampling is 
0.7 arcsec/pixel. The difference in velocity between the two components is about 250 km/s. }

\end{figure} 

%Figure 17
\begin{figure} 
%%\centerline{\psfig{figure=n2110-f17.cps,width=17cm,angle=270}} 

\figcaption{a) 2D rotational model  built with the stellar velocity curve and projected to the plane of the sky
at 42$^{\circ}$ inclination, the P.A of the line of nodes is 163.5$^{\circ}$ and the systemic velocity
is 2380 km/s. b) Residual velocity map resulting from the subtraction of the disk rotational model from the ionized
velocity gas traced by the [NII] emission line. The contours are plotted in steps of 10 km/s. Dashed blue contours
go from -80 to -10 km/s, and the full red contours from 10 to 100 km/s. }

\end{figure} 

%Figure 18
\begin{figure} 
%%\centerline{\psfig{figure=n2110-f18.ps,width=15cm,angle=270}} 

\figcaption{Unsharp masked image of NGC 2110 obtained dividing the HST+WFPC2 image through the filter F606W 
by a median filtered version of the same image (gray scale). 
The HI absorption cloud detected by Gallimore et al (1999)
is plotted as an ellipse.}

\end{figure}


\clearpage

\begin{thebibliography}{}

\bibitem[]{} Adams, T.F. 1977, ApJS, 33, 19
\bibitem[]{} Athanassoula, R. 1992, MNRAS, 259, 345
\bibitem[]{} Arribas, S., Mediavilla, E., Garc\'\i a-Lorenzo, B., del Burgo, C. 1997, ApJ, 490, 227
\bibitem[]{} Arribas, S., et al. 1998, Proc. SPIE, 3355, 821
\bibitem[]{} Arribas, S., Colina, L. \& Clements, D., 2001, ApJ, 560, 160
\bibitem[]{} Bak, J., 2000, PhD thesis, Ohio University
\bibitem[]{} Bingham, R.G., Gellatly, D.W., Jenkins, C.R., \& Worswick, S.P. 1994, Proc. SPIE, 2198, 56
\bibitem[]{} Binney, J.B., \& Tremaine, S.T., 'Galactic Dynamics', 1987, Princeton University Press. Pinceton: New Jersey
\bibitem[]{} Bradt, H.V., Burke, B.F., Canizares, C.R., Greenfield, P.E., Kelly, R.L., McClintock, J.E., 
van Paradijs, J. \& Koski, A.T. 1978, ApJ, 226, L111
\bibitem[]{} Canalizo, G. \& Stockton, A. 2001, ApJ, 555, 719
\bibitem[]{} Cecil, G., Bland, J. \& Tully, R.B. 1990, ApJ, 355, 70
\bibitem[]{} Cid Fernandes, R. \& Terlevich, R. 1995, MNRAS, 272, 423
\bibitem[]{} Cid Fernandes, R., Heckman, T., Schmitt, H.R, Gonz\'alez Delgado, R.M \& Storchi-Bergmann, T. 2001, ApJ, 558, 81
\bibitem[]{} Colbert, E.J.M., Baum, S.A., Gallimore, J.F., O'Dea, C.P., Lehnert, M.D., Tsvetanov, Z.I., Mulchaey, J.S. \& Caganoff, S. 1996, ApJS, 105, 75
\bibitem[]{} Colina, L., Gonz\'alez Delgado, R.M., Mas-Hesse, J.M. \& Leitherer, C. 2002, ApJ, submitted
\bibitem[]{} Colina, L., Arribas, S. \& Borne, K.D, 1999, ApJ, 527, L13
\bibitem[]{} de Vaucouleurs, G., de Vaucouleurs, A., Corwin, JR. H.G, Buta, R.J., Paturel, G. \& Fouqu\'e, P. 1991, 
Third Reference Catalogue of Bright Galaxies (Austin: Univ. of Texas) (RC3)
\bibitem[]{} Dahari, O. 1984, AJ, 89, 966
\bibitem[]{} De Robertis, M.M., Yee, H.K.C. \& Hayhoe, K. 1998, ApJ, 496, 93
\bibitem[]{} Ferruit, P., Wilson, A.S., Whittle, M., Simpson, C., Mulchaey, J.S. \& Ferland, G.J. 1999, ApJ, 523, 147
\bibitem[]{} Gallimore, J.F, Baum, S.A., O'Dea, C.P., Pedlar, A. \& Brinks, E. 1999, ApJ, 524, 684
\bibitem[]{} Garc\'\i a-Lorenzo, B., Mediavilla, E., Arribas, S. \& del Burgos, C., 1997, ApJ, 483, L99
\bibitem[]{} Garc\'\i a-Lorenzo, B., Mediavilla, E., \& Arribas, S. 1999, ApJ, 518, 190
\bibitem[]{} Gonz\'alez Delgado, R.M. 2001, in {\it Issues in Unification of AGNs}, in press (astro-ph/0109505)
\bibitem[]{} Gonz\'alez Delgado, R.M., Heckman, T. , Leitherer, C., Meurer, G., Krolik, J., Wilson, A.S., Kinney, A. \& Koratkar, A. 1998, ApJ, 505, 174
\bibitem[]{} Gonz\'alez Delgado, R.M., Heckman, T. \& Leitherer, C. 2001, ApJ, 546, 845
\bibitem[]{} Haynes, M.P., Jore, K.P., Barrett, E.A., Broeils, A.H., Murray, B.M. 2000, AJ, 120, 703
\bibitem[]{} Haynes, M.P., in 'Gas and galaxy Evolution', 2001, ASP Conf. vol. 240, p. 257
\bibitem[]{} Heckman, T.M., Gonz\'alez-Delgado, R., Leitherer, C., Meurer, G.R., Krolik, J., Kinney, A., Koratkar, A., \& Wilson, A.S. 1997, ApJ, 482, 114
\bibitem[]{} Hunt, L., Malkan, M.A., Salvati, M., Mandolesi, N., Palazzi, E, \& Wade, R. 1997, ApJS, 108, 229
\bibitem[]{} Knapen, J.H., Shlosman, I. \& Peletier, R. F. 2000, ApJ,529, 93
\bibitem[]{} Knop, R.A., Armus, L., Matthews, K., Murphy, T.W. \& Soifer, B.T. 2001, ApJ, 122, 764
\bibitem[]{} Krolik, J.H. \& Begelman, M.C. 1986, ApJ 308, L55
\bibitem[]{} Lehnert, M.D. \& Heckman, T.M. 1996, ApJ, 462, 651
\bibitem[]{} Lutz, D. \& Tacconi, L.J. 1999, in {\it Ultraluminous galaxies: monsters or babies}, ed. D. Lutz 
\& L.J. Tacconi, Astrophysic \& Space Science, 266
\bibitem[]{} Maiolino, R., Risaliti, G. \& Salvati, M. 1998, A\&A, 341, L35
\bibitem[]{} Malkan, M.A., Varoujan, G. \& Raymond, T. 1998, ApJS, 117, 25
\bibitem[]{} M\'arquez, I., Durret, F., Masegosa, J., Moles, M., Gonz\'alez Delgado, R.M., Marrero, I., Maza, J., P\'erez, E. \& Roth, M. 2000, A\&A, 360, 431
\bibitem[]{} Martini, P. \& Pogge, R.W. 1999, AJ, 118, 2646
\bibitem[]{} McClintock, J.E., van Paradijs, J., Remilliard, R.A., Canizares, C.R., Koski, A.T. \& Veron, P. 1979, ApJ, 233, 809
\bibitem[]{} Mihos, J.C., \& Hernquist, L. 1996, ApJ, 464, 641 
\bibitem[]{} Moles, M., M\'arquez, I. \& P\'erez, E. 1995, ApJ, 438, 604
\bibitem[]{} Mulchaey, J.S. \& Regan, M. 1997, ApJ, 482, L135
\bibitem[]{} Mulchaey, J.S. \& Regan, M. 1999, AJ, 117, 2676
\bibitem[]{} Mulchaey, J.S., Wilson, A.S., Bower, G., Heckman, T.M., Krolik, J.H. \& Miley, G.K. 1994, ApJ, 433, 625
\bibitem[]{} Mundell, C.G., Wilson, A.S., Ulvestad, J.S., \& Roy, A.L. 2000, ApJ, 529, 816
\bibitem[]{} Nagar, N.M., Wilson, A.S., Mulchaey, J.S. \& Gallimore, J.F. 1999, ApJS, 120, 209
\bibitem[]{} Nelson, C., \& Whittle, M. 1995, ApJSS, 99, 67
\bibitem[]{} Noguchi, M. 1988, A\&A, 203, 259
\bibitem[]{} Pogge, R.W. 1989, ApJ, 345, 730
\bibitem[]{} Pogge, R.W. \& Martini, P. 2002, ApJ, 569, 624
\bibitem[]{} Pringle, J.E., Antonucci, R.R.J., Clarke, C.J., Kinney, A.L., Schmitt, H.R. \& Ulvestad, J.S. 1999, ApJ, 526, L9
\bibitem[]{} Quillen, A.C., Alonso-Herrero, A., Rieke, M.J., Falcke, H. \& Rieke, G.H. 1999, ApJ, 525, 685
\bibitem[]{} Rafanelli, P., Violato, M. \& Baruffolo, A. 1995, AJ, 109, 1546
\bibitem[]{} Regan, M.W. \& Mulchaey, J.S. 1999, AJ, 117, 2676
\bibitem[]{} Sanders, D.B., Soifer, B.T., Elias, J.H., Neugebauer, G. \& Matthews, K. 1988, ApJ, 328, L35
\bibitem[]{} Shlosman, I., Begelman, M.C. \& Frank, J. 1990, Nature, 345, 679
\bibitem[]{} Shuder, J.M. 1980, ApJ, 240, 32
\bibitem[]{} Simien, F. \& Prugniel, Ph. 2000, A\&AS, 145, 263
\bibitem[]{} Storchi-Bergmann, T., Winge, C., Ward, M.J. \& Wilson, A.S. 1999, MNRAS, 304, 35 
\bibitem[]{} Strauss, M.A., Huchra, J.P., Davis, M., Yahil, A., Fisher, K.B. \& Tonry, J. 1992, ApJS, 83, 29
\bibitem[]{} Taniguchi, Y. 1999, ApJ, 524, L65
\bibitem[]{} Taylor, D., Dyson, J.E., \& Axon, D.J. 1992, MNRAS, 255, 351
\bibitem[]{} Tonry, J. \& Davis, M. 1979, AJ, 84, 1511
\bibitem[]{} Ulvestad J.S. \& Wilson, A.S. 1983, ApJ, 264, L7
\bibitem[]{} Veilleux, S. 2001, in {\it Starbursts Near and Far}, eds. Tacconi, L. \& Lutz, D. (Springer-Verlag), SSP 88, 88
\bibitem[]{} Veilleux, S., Shopbell, P.L. \& Miller, S.T. 2001, ApJ, 121, 198
\bibitem[]{} Weaver, K.A., Mushotzsky, R.F., Serlemitsos, P.J., Wilson, A.S., Elvis, M. \& Briel, U. 1995, ApJ, 442, 597
\bibitem[]{} Whittle, M. 1985, MNRAS, 213, 1 
\bibitem[]{} Wilkins, T.N. \& Axon, D.J., 1991, TWODSPEC, Starlink User Note, No. 16.
\bibitem[]{} Wilson, A.S. \& Baldwin, J.A., 1985, ApJ, 289, 124
\bibitem[]{} Wilson, A.S., Baldwin, J.A. \& Ulvestad, J.S. 1985, ApJ, 291, 627
\bibitem[]{} Zaristky, D., Smith, R., Frenk, C. \& White, S.D.M. 1997, ApJ, 478, 39



\end{thebibliography}
\end{document}